\documentclass[prl,aps,twocolumn,floats,nofootinbib]{revtex4}
\usepackage{amsmath,amssymb,graphicx,psfrag}

\begin{document}
\renewcommand{\ni}{{\noindent}}
\newcommand{\dprime}{{\prime\prime}}
\newcommand{\be}{\begin{equation}}
\newcommand{\ee}{\end{equation}}
\newcommand{\bea}{\begin{eqnarray}} 
\newcommand{\eea}{\end{eqnarray}}
\newcommand{\la}{\langle}
\newcommand{\ra}{\rangle} 
\newcommand{\dg}{\dagger}
\newcommand\lbs{\left[}
\newcommand\rbs{\right]}
\newcommand\lbr{\left(}
\newcommand\rbr{\right)}
\newcommand\f{\frac}
\newcommand\e{\epsilon}
\newcommand\ua{\uparrow}
\newcommand\da{\downarrow}
\title{Many-body localization-delocalization transition in quantum Sherrington-Kirkpatrick model}
\author{Sudip Mukherjee$^{1,2}$, Sabyasachi Nag$^{2}$ and Arti Garg$^{2}$} 
\affiliation{$^{1}$ Department of Physics.  Barasat Government College, Barasat, Kolkata 700124, India. \\
$^{2}$ Condensed Matter Physics Division, Saha Institute of Nuclear Physics, 1/AF Bidhannagar, Kolkata 700 064, India}
\vspace{0.2cm}
\begin{abstract}
\vspace{0.3cm}
We analyze many-body localization (MBL) to delocalization transition in Sherrington-Kirkpatrick (SK) model of Ising spin glass (SG) in the presence of a transverse field $\Gamma$. Based on energy resolved analysis, which is of relevance for a closed quantum system, we show that the quantum SK model has many-body mobility edges separating MBL phase which is non-ergodic and non-thermal from the delocalized phase which is ergodic and thermal. The range of the delocalized regime increases with increase in the strength of $\Gamma$ and eventually for $\Gamma$ larger than $\Gamma_{CP}$ the entire many-body spectrum is delocalized. We show that the Renyi entropy is almost independent of the system size in the MBL phase, hinting towards an area law in this infinite range model while the delocalized phase shows volume law scaling of Renyi entropy. We further obtain spin glass transition curve in energy density $\epsilon$-$\Gamma$ plane from the collapse of eigenstate spin susceptibility. We demonstrate that in most of the parameter regime SG transition occurs close to the MBL transition indicating that the SG phase is non-ergodic and non-thermal while the paramagnetic phase is delocalized and thermal.  
\vspace{0.cm}
\end{abstract} 
%\pacs{72.15.Rn, 05.30.Fk,05.30.Rt}
\maketitle
\section{Introduction}

Many-body localization (MBL) has been a topic of intense research in condensed matter physics in last decade. 
 The question of immense interest is what happens to Anderson localization~\cite{Anderson} in the presence of interactions. Recently Basko et. al.~\cite{Basko} have established that Anderson localization can survive weak interactions in the perturbative regime while strong interactions can destroy localization, resulting in a MBL-delocalization transition. Recently much insight in the field has been obtained based on numerical analysis of interacting one dimensional models of spin-less fermions or spins with completely random disorder~\cite{Husein,Huse2010,Bardarson2015,Sirker,Mueller2016} as well as models where there is no randomness but have a quasi-periodic or aperiodic potential~\cite{Huse2013,Subroto,Sdsarma,garg}. In many of these cases MBL has also been realised experimentally in cold atom experiments~\cite{expt}.

The MBL phase is non-ergodic and challenges the basic foundations of quantum statistical physics~\cite{Huse,Altman} based on which we expect any interacting non-integrable system to be ergodic. In the MBL phase the system explores only an exponentially small fraction of the configuration space. As a result of this ergodicity breaking, local observables do not thermalize leading to violation of eigenstate thermalization hypothesis (ETH)~\cite{Deutsch,Srednicki,Rigol}. Signature of these are also clearly visible in the entanglement entropy of the many-body spectrum~\cite{Nayak,Huse2013, Sdsarma}. An infinite temperature MBL phase has also been shown to have an  extensive number of local integrals of motion~\cite{Abanin,Mueller} and hence holds similarity with integrable systems~\cite{Huse2014,shastry}. 

Non-ergodic nature of the MBL phase leads to long time memory of the initial state in local observables. In this respect, the MBL systems exhibit glassy dynamics and the question of interest is whether classical models of spin-glasses will undergo MBL-delocalization transition in their corresponding quantum versions.  Recently MBL has been studied in spin-glass systems~\cite{Bardarson2014,Lee,Laumann,Baldwin} focusing mainly on the  question whether the spin glass phase is MBL or can be delocalized as well? In disordered transverse field Ising model~\cite{Bardarson2014} an intermediate delocalized SG phase has been observed between MBL SG and the delocalized PM phase.

In this work we explore MBL-delocalization transition in the Sherrington-Kirkpatrick (SK) model, which is a paradigmatic model of Ising spin-glass~\cite{SK,Young_rev}. The simplest approach to make this model quantum is by adding a transverse field $\Gamma$. There have been numerous works on quantum SK model mainly focusing to get the quantum critical point and the associated exponents so far~\cite{QSK} but not in context of MBL. Recently quantum SK model has been studied in context of MBL ~\cite{Baldwin} where it has been proposed that the spin glass phase in quantum SK model is non-ergodic which violates ETH but is not MBL either. In the presence of random longitudinal field this model~\cite{Burin} was recently mapped to Anderson model on Bethe lattice and MBL transition was analyzed in the paramagnetic phase as a function of temperature to conclude that the PM phase is delocalized in the thermodynamic limit.  

 In the SK model, all Ising spins are coupled to each other with couplings themselves being random. How interactions in this model affect localization is hence an interesting issue to be explored. Further due to its infinite range of interaction among spins, studying MBL transition in SK model becomes even more important because of the Griffith effects in the MBL phase~\cite{Agarwal_rare}. Effects of rare ``localized'' regimes in the thermal phase and rare ``thermal'' regimes in the MBL phase have been well understood for systems with short-range interactions in one dimension and higher~\cite{Ponte}. But the situation with long-range interactions is not so clear. We show in this work that in the quantum SK model, we do get a fraction of many-body states localized for a broad range of parameters. The reason might be that in the quantum SK model the interactions themselves are random and help in stabilizing the MBL phase. 

In this work, we study quantum SK model using full exact diagonalization focusing on the energy resolved analysis of various physical quantities relevant for the MBL to delocalization transition as well as for the spin-glass transition. We demonstrate that in most of the parameter regime, the SG phase is MBL which is non-ergodic and violates ETH while the paramagnetic phase is delocalized ergodic phase which obeys ETH. The delocalized PM phase in energy space becomes broader with increase in $\Gamma$ such that for $\Gamma > \Gamma_{CP}$ the entire spectrum is delocalized. 
We do not see clear signatures of any non-ergodic thermal phase or a spin glass phase which is delocalized.  
The rest of the paper is organized as follows. In the first section we describe the model, followed up by the details of various quantities relevant for  analysis of MBL-delocalization transition and spin glass transition, calculated using numerical exact diagonalization in section of results. We present the detailed phase diagram based on these quantities and conclude.

\section{Model}
The model we study is the quantum SK model of Ising SG, described by the following Hamiltonian for the system of N spins:
\bea
H=-\sum_{i<j}J_{ij}\sigma_i^z\sigma_j^z - \Gamma\sum_i \sigma_i^x
\label{model}
\eea
Here $\sigma_i^z$, $\sigma_i^x$ are the z and x components of Pauli spin matrices respectively and $\Gamma$ denotes the transverse field. 
The spin-spin couplings ($J_{ij}$) are distributed following Gaussian distribution $\rho(J_{ij}) = \big(\frac{N}{2\pi J^2}\big)^\frac{1}{2}\exp\big(\frac{-N J^2_{ij}}{2J}\big)$, where the mean and the variance of the distribution are zero and $J/\sqrt{N}$ respectively. Note that the Hamiltonian~(\ref{model}) has a global $\mathcal{Z}_2$ symmetry. 
For $\Gamma=0$, it is a classical mean-field $N$ spin Ising model in which there are quenched random interactions between all pairs of spins.
 It has an ordered SG phase below the critical temperature $T_g = 1$~\cite{Campbell}.
At $T=0$, the ground state of quantum SK model undergoes a continuous quantum phase transition from spin-glass phase to paramagnetic (PM) phase at $\Gamma_{CP}=1.5J$~\cite{QSK}.

Interesting feature about this model is the long range interaction between spins and the coupled source of randomness and interaction. Hence it provides an interesting playground to explore the possibility of MBL phase and analyze MBL-delocalization phase transition accompanied by the SG transition.
\begin{figure}[h!]
\begin{center}
%\hskip-0.1cm
\includegraphics[width=2.35in,angle=-90]{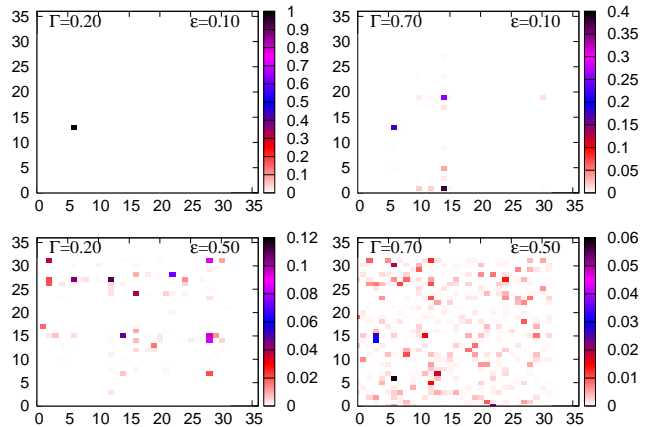}
\caption{Density plots of $|\Psi_n(i)|^2$ where $i$ the basis state label has been decomposed into $(i_A,i_B)$ such that $i_A$ is label for first $N/2$ sites (shown on the x-axis) and $i_B$ (shown on the y axis) is the label for right $N/2$ sites with $i=i_B+ 2^{N/2}i_A$. Wavefunction for $\epsilon=0.1$ for $\Gamma=0.2$ is clearly localized while that for $\epsilon=0.5$ and $\Gamma=0.7$ is clearly extended. These plots are for $N=10$ for a given disorder configuration.}
\label{wvfn}
\end{center}
\end{figure}
\section{Results}
The results described in the following sections are obtained by solving the model in Eq.~\ref{model} using exact diagonalization on finite size systems. 
The phase diagram shown in Fig.~\ref{phase_diag} has been obtained on the basis of analysis of various quantities namely, energy spacing statistics, normalized participation ratio, Shannon entropy, Renyi entropy and ETH. We also analyzed spin-glass susceptibility in order to find energy resolved spin-glass to PM phase transition. Below we describe our results for each of these quantities one by one. 

\subsection{Localization in Fock space}
Since MBL is the localization in Fock space, to decide whether a many-body state is localized or not, we first look at the eigenfunctions $\Psi_n(i)$ of the Hamiltonian in Eq.~\ref{model} with eigenvalues $E_n$ in the Fock space. Fig. [\ref{wvfn}] shows the density plot of $|\Psi_n(i)|^2$ in the configuration space with $i\in[0,2^N-1]$. We decompose each configuration label $i$ into that of two equal subsystems as $(i_A,i_B)$ such that $i=i_B+2^{N/2}i_A$ with both $i_A$ and $i_B$ having $2^{N/2}$ values which are used as x and y axis of the square configuration space in Fig.~[\ref{wvfn}]. It is clear that the low energy states e.g. with $\epsilon=0.1$, where $\epsilon =  \frac{E-E_{min}}{E_{max} - E_{min}}$ is the normalized energy density, are highly localized in the configuration space for small values of $\Gamma$. As $\Gamma$ increases these states at the bottom of spectrum gets slightly more delocalized. States in the middle of the spectrum corresponding to $\epsilon=0.5$ are already delocalized for $\Gamma=0.2$ and become highly extended with almost equal weight for all basis states for large values of $\Gamma=0.7$ as shown in bottom right panel of Fig.[\ref{wvfn}]. 
 \begin{figure}[h!]
\begin{center}
%\hskip-0.1cm
\includegraphics[width=2.35in,angle=-90]{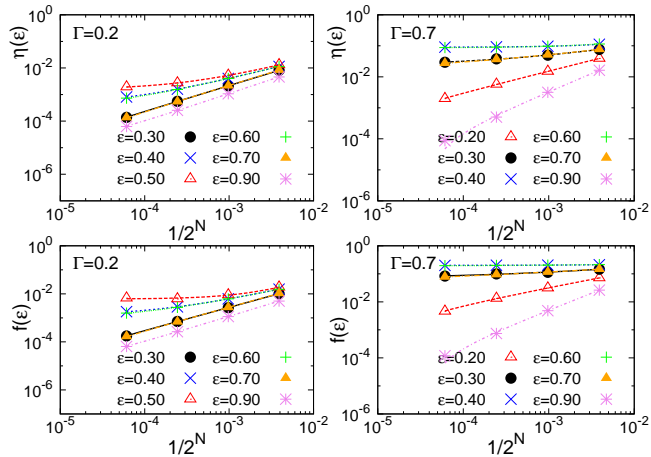}
\caption{Top panel: $\eta(\epsilon)$ vs $1/V_H$ for various values of $\epsilon$. The left panel shows results for $\Gamma=0.2$ and the right panel shows results for $\Gamma=0.7$ in units of $J$. Here for $\Gamma=0.2J$, $\eta(\epsilon) \sim b*V_H^{-c}$ for all values of $\epsilon$ except for a few states with $\epsilon \sim 0.5$ where  $\eta(\epsilon)\sim \eta_0+b*V_H^{-c}$ .
On increasing $\Gamma$ more many body states get delocalized as indicated in panel (b). The bottom panel shows results for $f(\epsilon)=\exp(S(\epsilon))/V_H$ obtained from Shannon entropy.}
\label{NPR}
\end{center}
\end{figure}

\begin{figure}[h!]
\begin{center}
\hskip-0.8cm
\vskip-1.3cm
\includegraphics[width=2.5in,angle=-90]{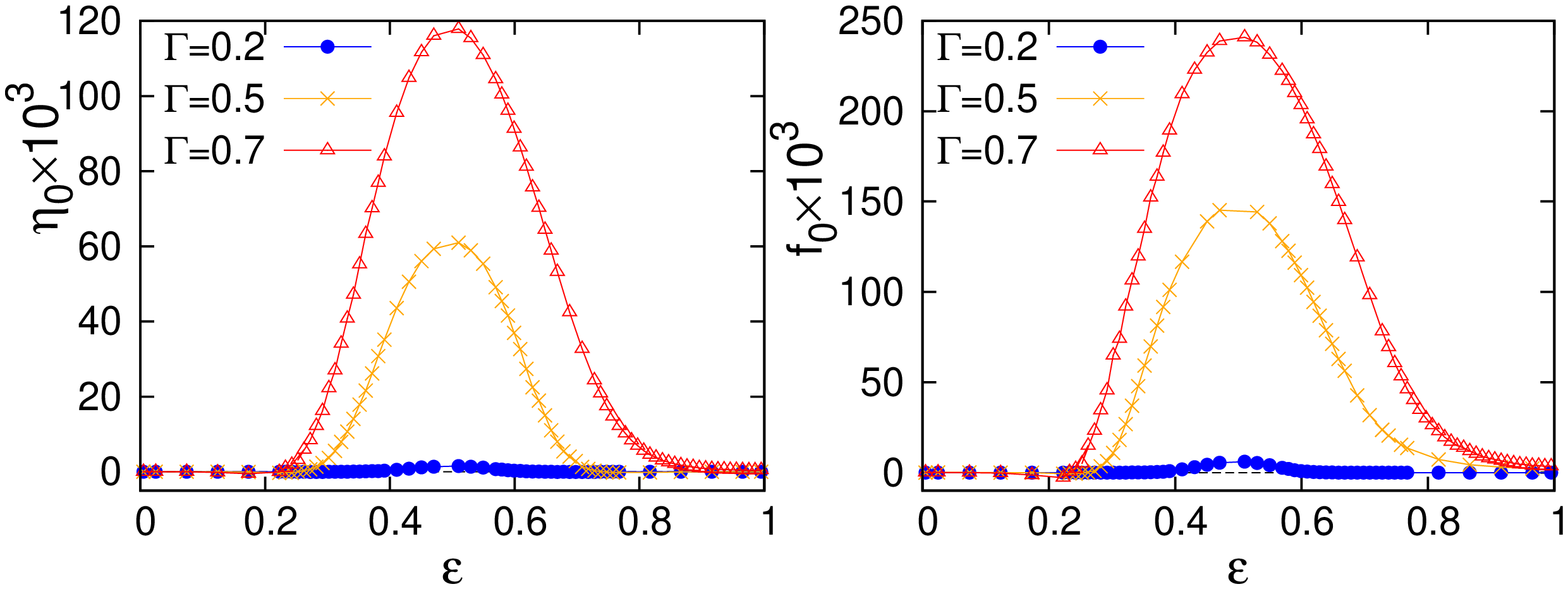}
\vskip-3cm
\includegraphics[width=2.5in,angle=-90]{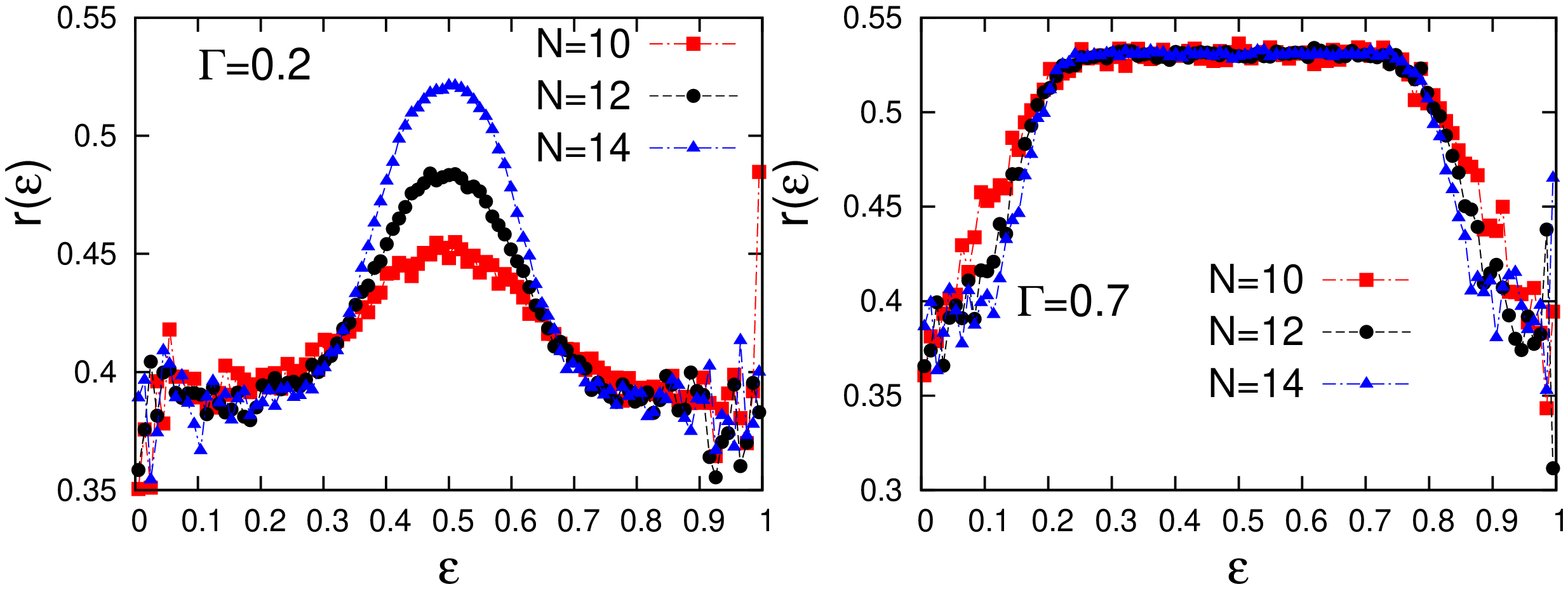}
\vskip-1.5cm
\caption{Top Panel: Extrapolated value of NPR in the thermodynamic limit $\eta_0$ vs $\epsilon$ for a few values of $\Gamma$. The right top panel shows the extrapolated value of function $f_0$ obtained in the thermodynamic limit from the Shannon entropy vs $\epsilon$. Clearly the delocalized regime increases as $\Gamma$ increases. Bottom Panel: Ratio of successive gaps $r(\epsilon)$ vs $\epsilon$ for $\Gamma=0.2$ and $\Gamma=0.7$ for different system sizes. For $\epsilon < \epsilon_1$ and $\epsilon >\epsilon_2$, $r(\epsilon)$ is close to its value for PS and does not increase with $N$. But for intermediate energy values $r(\epsilon)$ increases with $N$ and approaches the value for WDS.}
\vskip-0.5cm
\label{r}
\end{center}
\end{figure}

{\bf{ Normalized Participation Ratio:}} To quantify the amount of localization, we calculate the normalized participation ratio (NPR) which is defined as
$\eta(\epsilon) = \frac{1}{\langle \sum_{i,n} |\Psi_n(i)|^4 \delta(E-E_n) \rangle_C V_H}$, 
where $V_H=2^N$ is the volume of the Fock space for system of $N$ sites, and $\langle \rangle _C$ indicates the configuration averaging.  We replace the delta function by a box distribution of finite width $d E$ around dimension-less energy $\epsilon$ to obtain $\eta(\epsilon)$. $\eta(\epsilon)$ represents the fraction of configuration space participating in a many-body state of energy $\epsilon$. For delocalized states, $\eta(\epsilon)$ is of order $O(1)$ while for states showing MBL, $\eta(\epsilon)$ decreases with increase in the system size and vanishes in the thermodynamic limit.
Fig.~\ref{NPR} shows scaling of $\eta(\epsilon)$ w.r.t $1/V_H$. For very low and high energy states, $\eta(\epsilon)$ decreases as the system size increases going to zero in the thermodynamic limit ($\eta(\epsilon)\sim bV_H^{-c}$) indicating localized nature of these many body states. For states in the middle of the band $\eta(\epsilon) \sim \eta_0+b(1/V_H)^c$ with finite value $\eta_0$ in the thermodynamic limit indicating the ergodic nature of these states. The top panel of Fig.~[\ref{r}] shows extrapolated values $\eta_0$ vs $\epsilon$ which determines the transition point $\epsilon_{1,2}$ such that for $\epsilon<\epsilon_1$ and $\epsilon > \epsilon_2$, $\eta_0 \sim 0$ while in the intermediate energy states $\eta_0$ is finite. 

{\bf{Shannon Entropy}}: For conventional single particle Anderson localization problem, participation ratio is associated with the spread of a single particle state in real space, but in the abstract Fock space there is no clear concept of distance between two points of the Fock space. Hence we used another measure to check localization in Fock space, which is the Shannon entropy for every eigenstate  $S(E_n)=-\sum_{i=1}^{V_H}|\Psi_n(i)|^2\ln|\Psi_n(i)|^2$. Clearly for a many body state which gets contribution from all the basis states in the Fock space (and is normalized) $S(E_n)\sim \ln(V_H)$. Thus $f(\epsilon)=\exp(S(\epsilon))/V_H$ (obtained by averaging over all eigenstates in a small energy window around $\epsilon$) is of  order unity $f(\epsilon) \sim 1$ for a delocalized state while for a localized state which gets significant contribution only from some of the basis states , say $N_l$, in the Fock space, then $f(\epsilon) \sim N_l/V_H$ vanishing to zero in the thermodynamic limit.
Bottom panel of Fig.~\ref{NPR} shows scaling of $f(\epsilon)$ w.r.t $1/V_H$. Interestingly, $f(\epsilon)$ has behavior very similar to $\eta(\epsilon)$ which further confirms our observation that top and bottom many-body states remain localized in the presence of finite $\Gamma$ though the delocalized regime expands with increase in $\Gamma$. This is clearly indicated in the top right panel of Fig.~[\ref{r}] which shows extrapolated value $f_0$ in the thermodynamic limit as a function of $\epsilon$ for a few $\Gamma$ values. Note that the transition points obtained from Shannon entropy coincide with $\epsilon_{1,2}$ obtained from the NPR.  

\subsection{Energy level spacing statistics}
A convenient measure to differentiate between the localized and extended states is based on study of spectral statistics using tools from random matrix theory~\cite{Mehta}. The distribution of energy level spacings is expected to follow Poisson statistics (PS) for many body localized phase while it follows Wigner-Dyson statistics (WDS) for the ergodic phase.  Following ~\cite{Husein}, we calculate the ratio of successive gaps in energy levels $r_n=\frac{min(\delta_n,\delta_{n+1})}{max(\delta_n,\delta_{n+1})}$ with $\delta_n=E_{n+1}-E_{n}$ at a given Eigen energy $E_n$ of the Hamiltonian in Eq.~\ref{model} to discriminate between the two phases. For a Poissonian distribution, the disorder averaged value of $r_n$ is $2ln2-1\approx 0.386$; while for the Wigner surmise of Gaussian orthogonal ensemble (GOE) mean value of $r_N$ is approximately $0.5295$. 

Fig.~\ref{r} shows the plot of $r(\epsilon)$ vs $\epsilon$ for various system sizes. This data is obtained from $r_n$ by averaging over 200-500 independent configurations. For smaller $\Gamma$ values, as shown in the left panel, for many-body states at the top and the bottom of the spectrum, $r$ for all the system sizes studied is close to its value expected for Poissonian distribution indicating their localized nature. On the other hand, for $\epsilon \sim 0.5$, $r$ increases with the system size approaching to the average value for the WD distribution. From system size dependence of $r$, one can obtain the transition points in energy density for a given $\Gamma$ which are very close to those obtained from NPR and Shannon entropy. For larger values of $\Gamma\sim J$, for most the many body states, $r$ is close to its value for the WD distribution except for a few states on the edges of the spectra where the states are localized. 
\begin{figure}[h!]
\begin{center}
\includegraphics[width=2.35in,angle=-90]{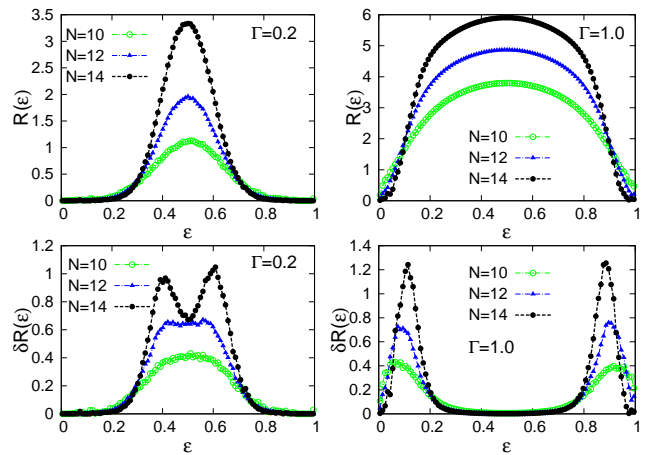}
\caption{Top panel shows $R(\epsilon)$ vs $\epsilon$ for various values of $N$. For $\epsilon< \tilde{\epsilon}_1$ and $\epsilon > \tilde{\epsilon}_2$, $R(\epsilon)$ is same for all $N$ values but for the intermediate states $R(\epsilon)$ increases with $N$ indicating their ergodic nature. The ergodic regime expands with increase in $\Gamma$ as shown in the right panel. The bottom panel shows the variance $\delta_R(\epsilon)$ of EE as a function of $\epsilon$. $\delta_R(\epsilon)$ shows two clear peaks, indicating localization to delocalization transition in the many-body spectrum. 
 }
\label{Renyi}
\end{center}
\end{figure}
\subsection{Entanglement Entropy}
Entanglement entropy (EE) characterizes how information spreads from one part of the system to another and is a useful tool to distinguish between the ergodic and many-body localized phases. Though for short range interacting systems, EE for the ground state obeys area law while higher excited states show volume law scaling $N^{d-1}$~\cite{area_law}, in the MBL phase even higher excited states adhere to the area law scaling of EE provided the range of interactions is short ~\cite{Nayak,Huse2013,Sdsarma}. However, there is currently no general and rigorous understanding of when area laws do or do not survive in the presence of long-range interactions~\cite{LR_area1,LR_area2}, which is the case of model in Eq.~(\ref{model}). Though in many of the cases, for power law interactions $V(r) \sim r^{-\alpha}$ with $\alpha > d+2$ ground state continues to show area law of EE, for long range cases with $\alpha < d$, logarithmic dependence on system size is seen~\cite{LR_area2}. SK model being the infinite range model should be close to $\alpha < d$ case though due to randomness of the interaction itself the situation is more complicated here. 

In order to develop understanding of the situation for quantum SK model, we evaluate the bipartite entanglement entropy. We divide the lattice into two subsystems A and B of sites $N/2$ and calculate the energy resolved Renyi entropy $R(E_n) = -log[Tr_A \rho_A(E_n)^2]$ where $\rho_A$ is the reduced density matrix obtained by integrating the total density matrix $\rho_{total}(E_n) = |\Psi_n\ra \la \Psi_n|$ over the degree of freedom of subsystem B.  
Top panel of Fig.~\ref{Renyi} shows $R(\epsilon)$ which is obtained from $R(E_n)$ by binning over an energy bin $dE$ around energy $\epsilon$ and averaging over 200-500 configurations. For $\epsilon < \tilde{\epsilon}_1$, $R(\epsilon)$ is same for various system sizes (within numerical error) indicating
 the localized nature of these states. It is interesting to observe area law behaviour of EE in this infinite range system. Same is true for many body states at the top of the spectrum with $\epsilon > \tilde{\epsilon}_2$, while for many-body states in the middle of the spectrum $\tilde{\epsilon}_1< \epsilon< \tilde{\epsilon}_2$, $R(\epsilon)$ increases with the system size following the volume law $R \sim N$. As $\Gamma$ increases, the width of intermediate states, which show volume law scaling of Renyi entropy, increases. We would like to emphasize that this picture is qualitatively consistent with what we obtained from the analysis of NPR and Shannon entropy though values of $\tilde{\epsilon}_{1,2}$ are little off from $\epsilon_{1,2}$ obtained earlier as shown in the Fig.~\ref{phase_diag}. Note that within the system sizes studied, there is no systematic trend in values $\epsilon_{1,2}$ w.r.t $\tilde{\epsilon}_{1,2}$, hence we believe that there is no non-ergodic phase which shows volume law scaling of EE in this system.  

The bottom panel of Fig.~\ref{Renyi} shows the variance of EE  $\delta_R(\epsilon)=\la R(\epsilon)^2\ra-\la R(\epsilon) \ra^2$ for various values of $\Gamma$ and $N$. In the thermodynamic limit $\delta_R$ should be zero deep inside the localized and delocalized phases but at the transition point it diverges due to contribution from both the extended and the localized states~\cite{Bardarson2014} which is reflected as a peak in finite size calculations. Notice that the peaks get sharper with increase in the system size as expected. Our data shows two clear peaks in $\delta_R(\epsilon)$ vs $\epsilon$ curve for a given $\Gamma$ indicating two transition points.  But these transition points are little off both from the  $\epsilon_{1,2}$ obtained from NPR analysis as well as from scaling of EE.

\begin{figure}[h!]
\begin{center}
\hskip-0.4cm
\includegraphics[width=2.35in,angle=-90]{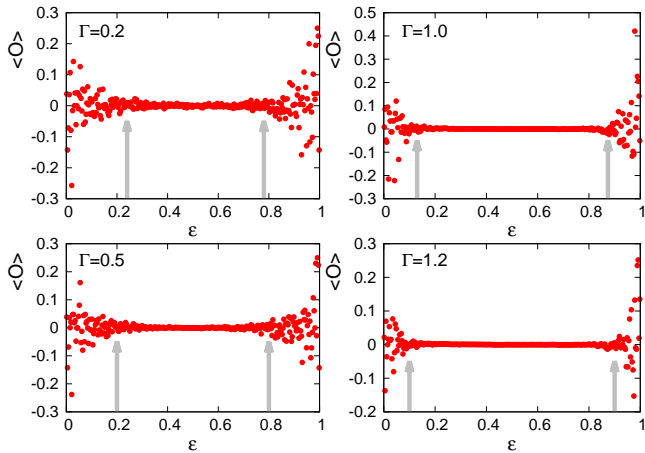}
\caption{$O(\epsilon)$ vs $\epsilon$ for a few values of $\Gamma$ which shows large fluctuations in its value for near by eigenstates for $\epsilon <\tilde{\tilde{\epsilon}}_1\sim \tilde{\epsilon}_1$ and $\epsilon>\tilde{\tilde{\epsilon}}_2\sim \tilde{\epsilon}_2$. As $\Gamma$ increases, width of non-thermal regime decreases while the middle thermal region increases in energy space.}
\label{ETH}
\end{center}
\end{figure}
\subsection{Eigenstate Thermalisation Hypothesis}
To check for the ETH in various parameter regimes we calculated expectation value of the $\sigma^z_i$ operator on subsystem A which has $N/2$ sites. We, therefore,  define $\hat{O} = \sum_{i=1}^{N/2} \hat{\sigma^z}_i$ and calculate its expectation value w.r.t all eigenstates. As shown in Fig.~\ref{ETH}, for $\epsilon < \tilde{\tilde{\epsilon}}_1$ and $\epsilon > \tilde{\tilde{\epsilon}}_2$ (shown by arrows), many-body system is not thermal showing large fluctuations in $\la O(\epsilon) \ra$ for nearby energy states  while for $\tilde{\tilde{\epsilon}}_1 < \epsilon< \tilde{\tilde{\epsilon}}_2$, system obeys ETH. We would like to emphasize that $\tilde{\tilde{\epsilon}}_{1,2} \sim \tilde{\epsilon}_{1,2}$ obtained from the scaling of Renyi entropy. Violation of ETH in the low and high energy part of the spectrum further confirms our observation that the EE in this regime follows area law.

\begin{figure}[h!]
\begin{center}
\hskip-0.4cm
\vskip-1cm
\includegraphics[width=2.5in,height=2.7in,angle=-90]{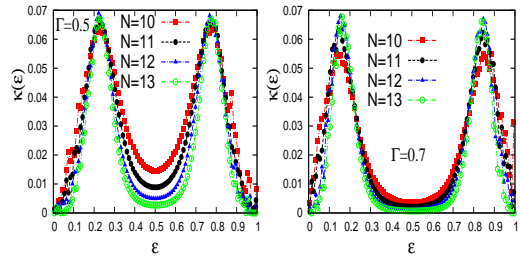}
\vskip-2cm
\caption{ Generalized PR obtained from off-diagonal matrix elements $\kappa(\epsilon)$ vs $\epsilon$ for a few values system sizes. $\kappa(\epsilon)$ shows two clear peaks indicating transition from non-ergodic to ergodic phases. Note that the peak positions are very close to the transition points $\tilde{\epsilon}_{1,2}$ obtained from the EE and the diagonal matrix elements $O(\epsilon)$.}
\label{ETH2}
\vskip-1cm
\end{center}
\end{figure}

We also looked at the off-diagonal matrix element $F(i)_{n,m} = \la\Psi_n|\sigma^z_i|\Psi_m\ra$ around an energy density $\epsilon$. Following~\cite{Baldwin} we defined a generalized participation ratio $\kappa(i,n)= \sum_{m\ne n} |F(i)_{n,m}|^4$ such that $|\epsilon_n-\epsilon_m| \le d\epsilon$, which is averaged over site index $i$ for a given configuration and then averaged over various disorder configurations to obtain $\kappa(\epsilon)$ at an energy density $\epsilon$. 
Fig.~\ref{ETH2} shows $\kappa(\epsilon)$ vs $\epsilon$ for a few system sizes. At the transition point $\kappa(\epsilon)$ diverges due to contribution from both thermal and non-thermal states. In a finite system size calculation, this divergence appears as peaks which separates the non-thermal states from the thermal ergodic states. Note that the peak positions in $\kappa$ vs $\epsilon$ curve are very close to $\tilde{\epsilon}_{1,2}$ obtained from diagonal elements $\la O \ra$ and from the scaling of EE.  In the delocalized ergodic phase, which obeys ETH, $\kappa(\epsilon)$ is supposed to decay exponentially with the system size~\cite{Baldwin} while in a non-ergodic MBL phase it should be of order $O(1)$. As shown in Fig.~\ref{ETH2}, for the energy states in the middle of the spectrum there is a clear suppression of $\kappa$ with the system size while on the non-ergodic side, $\kappa$ is almost independent of the system size~\cite{footnote}. 

\begin{figure}[h!]
\begin{center}
\hskip-1.cm
\vskip0.5cm
\includegraphics[width=2.75in,angle=0]{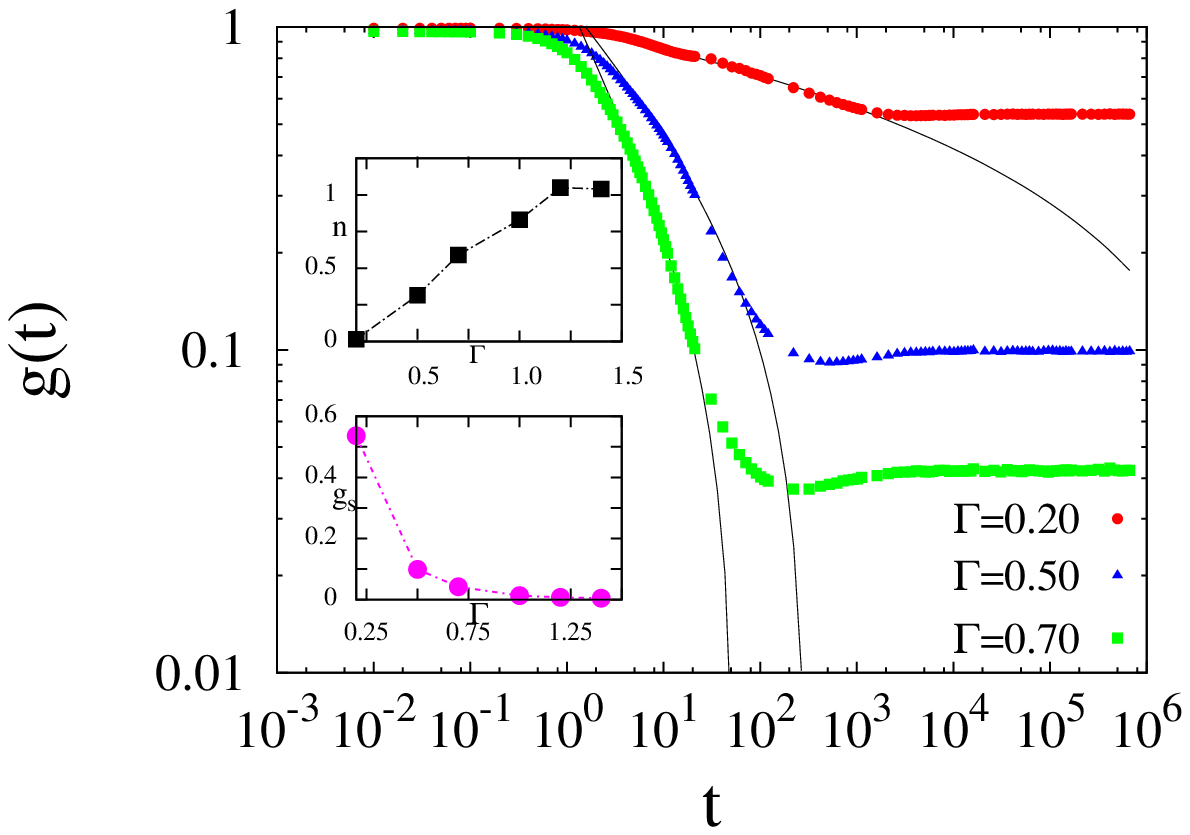}
\hskip3cm
\includegraphics[width=2.35in,height=2.2in,angle=-90]{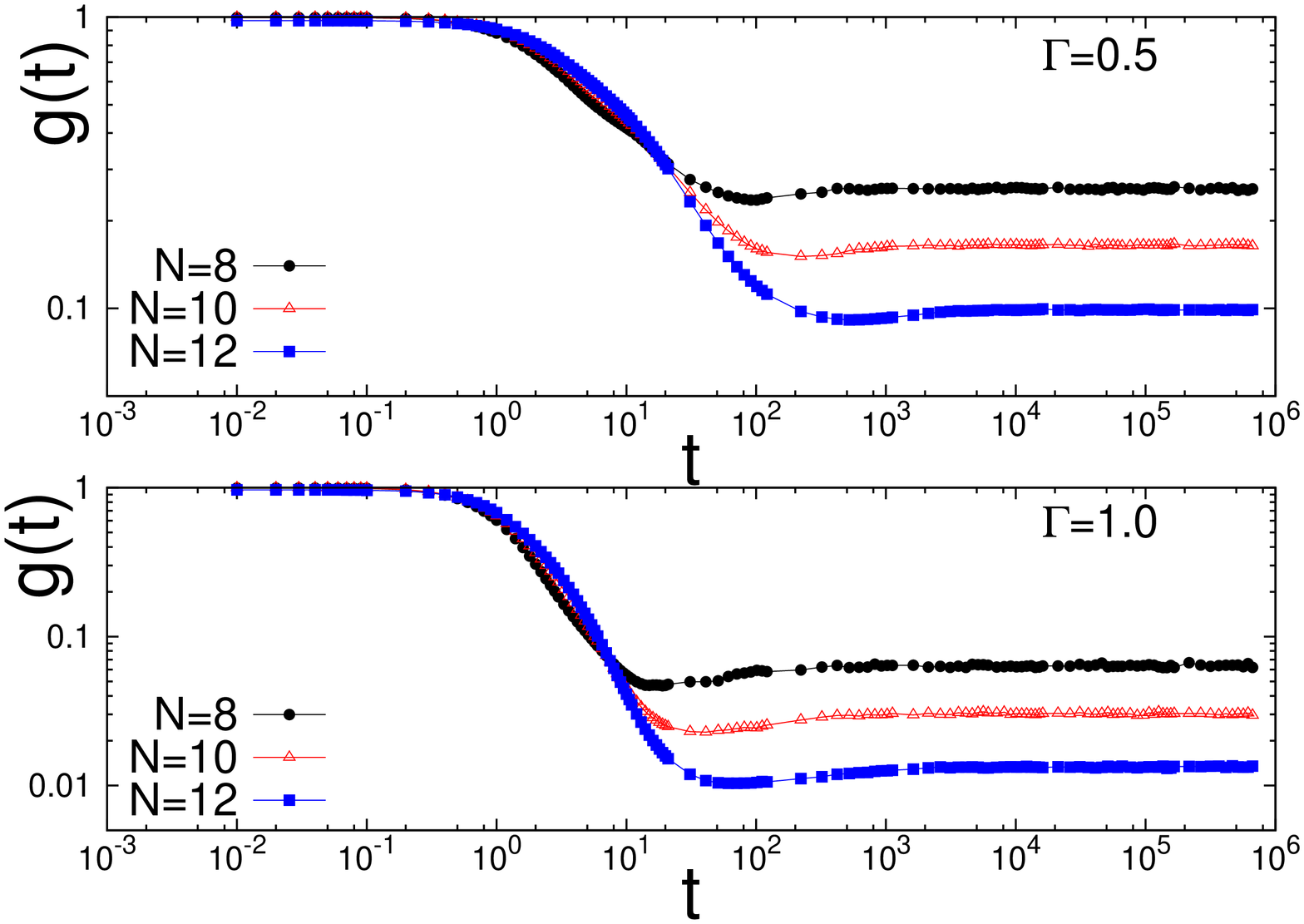}
\caption{Top panel: Auto-correlation function $g(t)$ for various values of $\Gamma$. For small values of $\Gamma$, $g(t)$ does not show significant decay with time, attaining a large value close to $1$ in the long time limit, indicating localized nature of the system. For larger values of $\Gamma$, $g(t)$ decreases rapidly with $t$ indicating its delocalized nature. Black lines show fit to the power law $g(t)\sim t^{-n}$. Insets show the exponent $n$ vs $\Gamma$ and the long time limit of the correlation function $g_s$. This data is for $N=12$. Middle and bottom panel shows system size dependence of $g(t)$ for $\Gamma=0.5$ and $1.0$.}
\label{gt}
\end{center}
\end{figure}

\subsection{Auto-correlation function}
We further calculate the auto-correlation function defined as 
\be
g(i,t)=\frac{1}{V_H}\sum_{n=1}^{V_H}\langle n|\sigma_i^z(t=0)\sigma_i^z(t)|n\rangle
\ee
where $|n\rangle$ is the $n^{th}$ eigenstate of the Hamiltonian~(\ref{model}). For a given disorder configuration, we average over all sites followed by averaging over various independent disorder configurations to get the average value of the auto-correlation function $g(t)$. In literature $g(t)$ is also known as the return probability~\cite{Subroto,Agarwal2,garg}. As shown in the top most panel of Fig.~\ref{gt}, for small values of $\Gamma$, $g(t)$ decays a bit in time and attains a value close to the saturation value of unity, indicating that if a spin at a site $i$, for example, was $\ua$ at initial time $t=0$, even after a long time $t$ it remains up. The system in this parameter regime has strong memory effect which is a signature of the MBL phase.
As shown in the inset of top panel of Fig.~(\ref{gt}), $g(t)\sim t^{-n}$ and the exponent is close to zero for very small values of $\Gamma$. For larger values of $\Gamma$, $g(t)$ decays faster indicated by larger values of the exponent $n$ and smaller values of $g$ in the long time limit, which we denote as $g_s$ and is shown in the bottom inset of Fig.~(\ref{gt}).  For intermediate values of $\Gamma \sim 0.7J$, the system is in the diffusive regime with $n\sim 0.5$. Still in the long time limit $g_s$ has a non-zero finite value though very small. Here one has to remember that $g(t)$ is obtained by summing contribution from all the eigenstates. Since not all the many-body states in this $\Gamma$ regime are delocalized (which are diffusive in this regime), there is a finite fraction of localized state as well, which make $g_s$ finite. Eventually for $\Gamma \Rightarrow \Gamma_{L,max}\sim 1.5J$, when the entire many-body spectrum gets delocalized, $n\Rightarrow 1$ and $g_s\Rightarrow 0$ indicating fully ballistic extended phase of the system.     
The system size dependence of $g(t)$ is shown in middle and bottom panel of Fig.~[\ref{gt}].

\begin{figure}[h!]
\begin{center}
\hskip-0.4cm
\includegraphics[width=2.325in,angle=0]{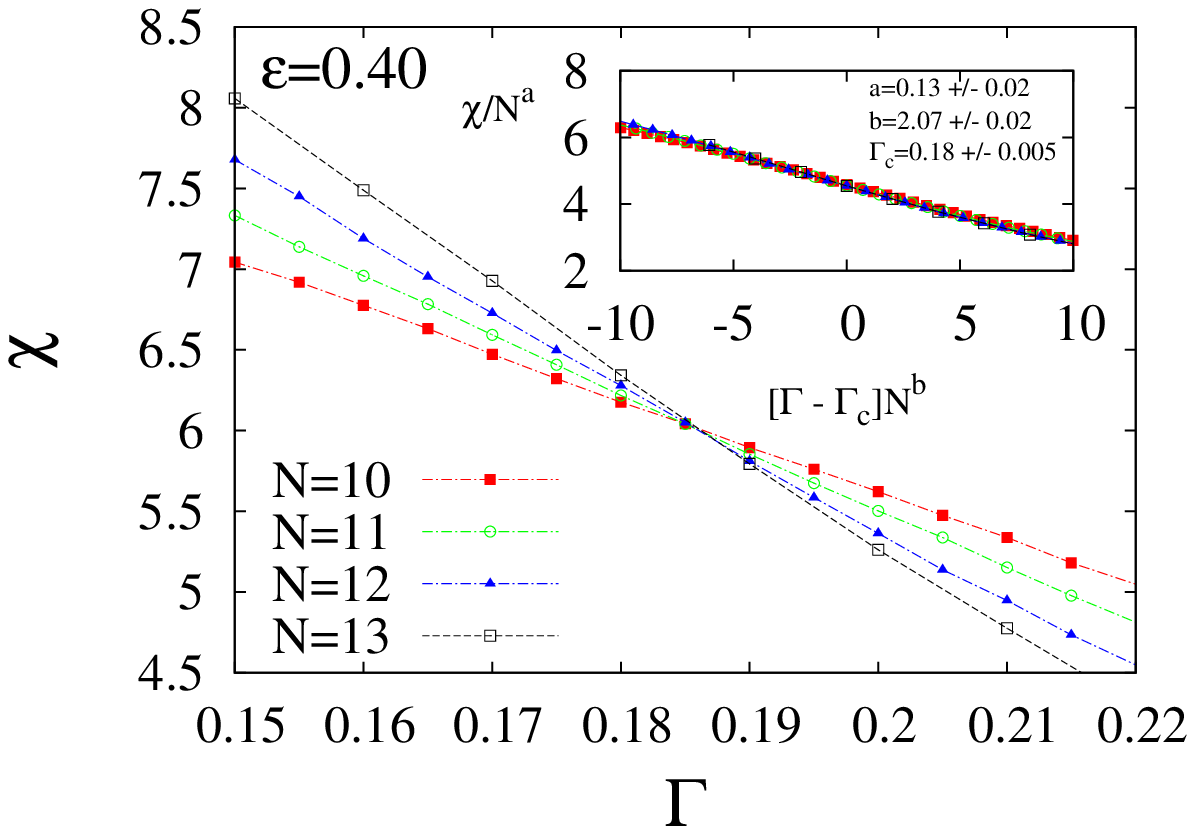}
\vskip-0.15cm
\includegraphics[width=2.35in,angle=0]{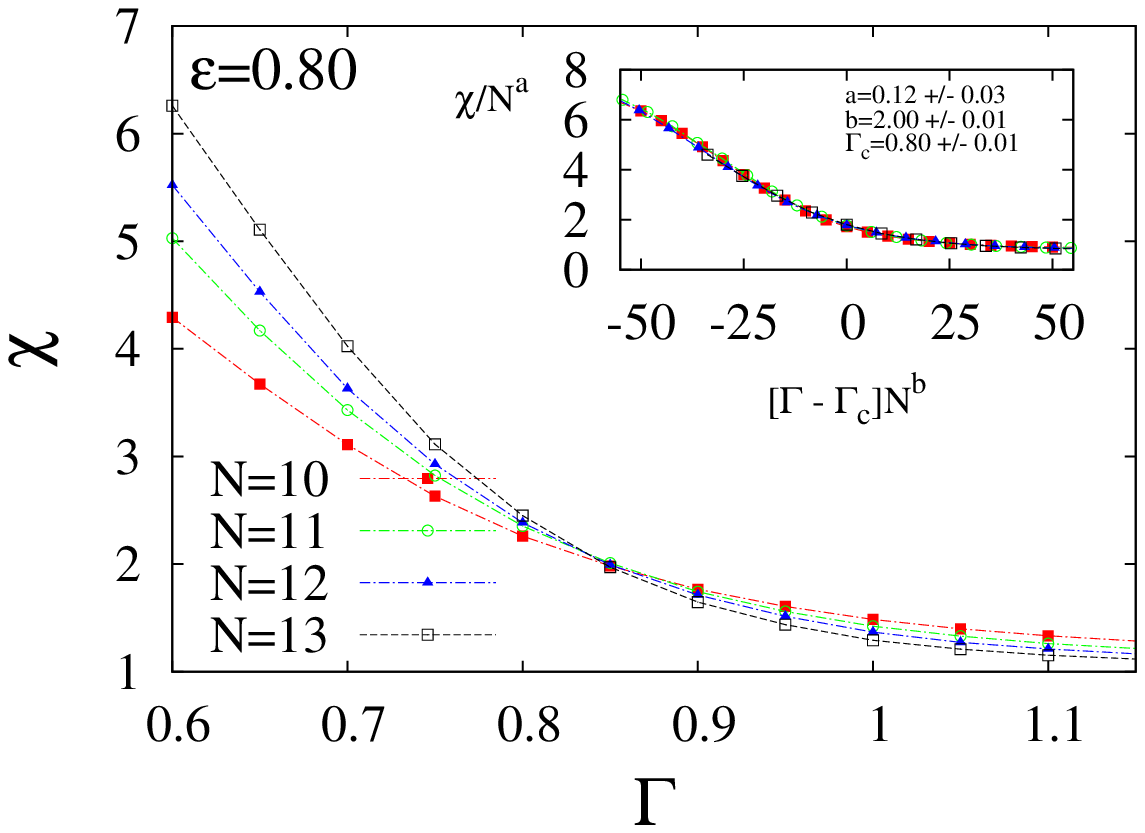}
\caption{ Spin-glass order parameter $\chi$ as function of $\Gamma$ for different system sizes $N$. Top panel shows the data for $\epsilon=0.4$ while the bottom panel is for $\epsilon=0.8$. In the SG phase $\chi$ is proportional to $N$. The inset shows the scaling collapse of the data.}
\label{chi_fig}
\end{center}
\end{figure}

\subsection{Spin susceptibility}
Various quantities described so far are tools to analyze MBL to delocalization transition. The system also has a SG phase accompanied by the broken $\mathcal{Z}_2$ symmetry protected by disorder. In order to identify the spin-glass to PM phase transition, we define the spin-glass order parameter as
\be
\chi_n = \frac{1}{N}\sum_{i,j=1}^N \langle n|\sigma_i^z\sigma_j^z|n\rangle^2
\label{chi}
\ee
Here $|n\rangle$ is the $n^{th}$ eigenstate of the Hamiltonian~(\ref{model}). Eventually $\chi(\epsilon)$ for certain value of $\epsilon$ is obtained by averaging over $\chi_n$ for all states within energy window $d \epsilon$ around $\epsilon$ for each disorder realization and then averaging over various independent disorder configurations.
In the SG phase, $\chi$ diverges in the thermodynamic limit $\chi_n \sim N$ while outside the SG phase $\chi_n \sim 1 $. 
Fig.~\ref{chi} shows numerical results for $\chi$ vs $\Gamma$ for two values of $\epsilon$ namely $0.4$ and $0.8$. 
For weak transverse field $\Gamma < \Gamma_c(\epsilon)$, when the system is in the SG phase, $\chi$ increases with the system size while for $\Gamma> \Gamma_c(\epsilon)$, $\chi$ decreases with the system size eventually approaching $1$. 
We obtain the spin-glass phase transition point $\Gamma_c(\epsilon)$ by performing a finite size scaling via the following equation:
\be
\chi(\epsilon,N,\Gamma) = N^af((\Gamma-\Gamma_c(\epsilon))*N^b)
\label{scaling}
\ee

By doing similar scaling analysis for various values of $\epsilon$ we get the transition curve $\Gamma_c(\epsilon)$ (shown in red in Fig.~\ref{phase_diag}) at which a continuous transition from spin-glass phase to the PM phase takes place. Maximum value of $\Gamma_c(\epsilon)=\Gamma_{c,max} \sim \Gamma_{L,max}$ is close to the value of quantum critical point $\Gamma_{CP}$ predicted from various earlier works on quantum SK model~\cite{QSK}.

\subsection{Phase Diagram}

\begin{figure}[h!]
\begin{center}
\hskip-0.4cm
\includegraphics[width=2.15in,angle=-90]{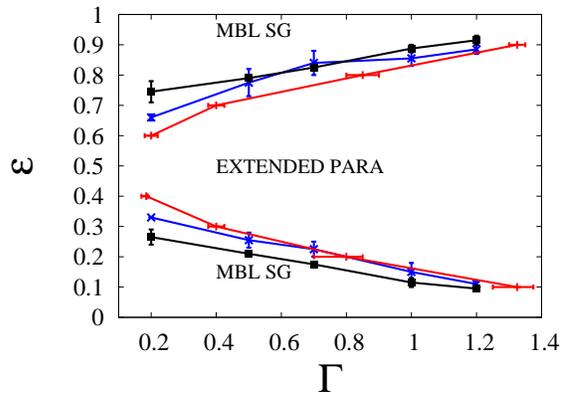}
\caption{Phase diagram of the quantum SK model (~\ref{model}) in the $\epsilon-\Gamma$ plane with $\epsilon$ being the relative energy density. The blue curve represents the energy resolved MBL to delocalization transition based on scaling of NPR, Shannon entropy and energy level spacing statistics. The black curve is based on Renyi entropy and ETH analysis  and the red curve, obtained on the basis of SG order parameter, represents the SG to PM phase transition. Within numerical precision, in most of the regime SG phase is MBL which is non-ergodic and violates ETH. 
 }
\label{phase_diag}
\end{center}
\end{figure}

Based on Renyi entropy scaling and ETH (both diagonal and off-diagonal matrix elements),  we obtain the thermal to athermal transition curve in $\epsilon-\Gamma$ plane shown by the black curve in Fig.~\ref{phase_diag}. Similarly, based on the scaling of NPR, Shannon entropy and energy level spacing statistics we obtain non-ergodic to ergodic transition curve in $\epsilon-\Gamma$ plane shown in blue in Fig.~\ref{phase_diag}. Note that for most of the regime with $\Gamma \ge 0.5$, the two transition curves are very close to each other indicating that non-ergodic states are non-thermal MBL states obeying area law for EE while all ergodic delocalized states are thermal obeying ETH and obey volume law for EE. For $\Gamma=0$ all states are localized and for any finite $\Gamma < \Gamma_{L,max}$, many-body states with $\epsilon < \epsilon_1$ and $\epsilon>\epsilon_2$ are localized while the intermediate states with $\epsilon_1<\epsilon<\epsilon_2$ are extended. The extent of delocalized spectrum increases with increase in $\Gamma$.  A many-body state with a given $\epsilon$ is localized for $\Gamma < \Gamma_L(\epsilon)$ and is delocalized for $\Gamma>\Gamma_{L}(\epsilon)$. The red curve shows the continuous spin glass transition curve $\Gamma_c(\epsilon)$ above which the system is PM and below which the system is in spin glass phase.

As clear from the phase diagram (Fig.~\ref{phase_diag}) the spin-glass transition point and the MBL transition point are very close to each other. In most of the parameter regime, spin-glass phase is non-ergodic and non thermal MBL phase and the PM phase is delocalized and thermal. Within the system sizes studied, it is not possible to claim anything too precisely very close to the transition points and hence possibility of a narrow ergodic SG phase near the transition points can not be ruled out either. 
  
Our numerical finding that the entire spin-glass phase is non ergodic is consistent with results of Baldwin et.al. ~\cite{Baldwin} where it was shown that that quantum SK model (which is a special case of quantum p-spin models with $p=2$) has non ergodic spin glass phase and there is no ergodic spin glass phase. Our numerical phase diagram is also consistent with analytical results of A. Burin~\cite{Burin} where MBL transition was studied as a function of temperature in the paramagnetic phase of quantum SK model to conclude that there is no MBL phase in the paramagnetic sector. 

\section{Conclusions}
In summary we have analyzed many body localization in a classical mean-field $N$ spin Ising model, namely the Sherrington-Kirkpatrick model, in which there are quenched random interactions between all pairs of spins, in the presence of a transverse field. Since the MBL systems have glassy dynamics in some sense, we wanted to explore whether the SK model, a paradigmatic model of classical SG, will exhibit many-body localization to delocalization transition when endowed with quantum mechanics. We demonstrated, based on the exact diagonalization study of the SK model in the presence of a transverse field $\Gamma$ that indeed this is true. For small values of $\Gamma$, many body states at top and bottom of the spectrum are localized in the Fock space and hence are non ergodic. These states also show violation of ETH while states in the middle of the band are delocalized in the Fock space, and hence are ergodic and thermal. We saw signatures of this energy resolved MBL-delocalization transition in normalized participation ratio, Shannon entropy, energy level spacings, EE and ETH. The delocalized regime increases with increase in $\Gamma$ and for $\Gamma \ge \Gamma_{L,max}$ all the many body states are delocalized. 

In systems with long range interactions, when and where EE obeys area law $R(E)\sim N^{d-1}$ is an unresolved issue. We demonstrated that for quantum SK model, which has infinite range interactions, for very low energy and high energy states which are non-ergodic and non-thermal, Renyi entropy is almost independent of the system size though for many body states in the intermediate energy range which are delocalised, Renyi entropy shows volume law scaling. 
We also calculated infinite temperature autocorrelation function which decays as $g(t)\sim t^{-n}$, with $n \sim 0$ for very small values of $\Gamma$ while $n \rightarrow 1$ for $\Gamma\ge 1.25$ indicating ballistic behaviour of the system while for intermediate values of $\Gamma$, $n \sim 0.5$. Long time value of $g(t)$ decreases with increase in $\Gamma$ becoming vanishingly small for $\Gamma \ge 1.25$. 
 
Generally due to rare region effects, there are questions about stability of the MBL phase in the presence of long range interaction in the system~\cite{Agarwal_rare}. We have demonstrated that in the quantum SK model, a fraction of many-body states are localized for a broad range of parameters. The reason might be that in the quantum SK model the interactions themselves are random and help in stabilizing the MBL phase. 

Further we explored the spin-glass transition by calculating the spin susceptibility $\chi(\epsilon)$ for each eigenstate and obtained the spin-glass to PM phase transition point by scaling collapse of the susceptibility data for various system sizes. Based on our numerical analysis, we conclude that SG to PM transition occurs very close to the MBL to delocalization transition such that the  spin glass phase is fully MBL while the paramagnetic phase is delocalized. Within numerical precision we have, we do not see any clear signatures of non-ergodic thermal phase or a SG phase which is delocalized.

In summary our work provides a detailed and complete analysis of the entire many body spectrum of quantum SK model and demonstrates that the spin glass phase is non-ergodic and non thermal while the paramagnetic phase is delocalized obeying ETH. We hope our work will help in understanding other aspects of the SK model related to the physics of quantum annealing and optimization problems.
\section{Acknowledgements}
We would like acknowledge Chris Baldwin, A. Pal, C. Laumann, and A. Scardicchio for detailed comments on the manuscript. We would also like to acknowledge A. Burin for clarifications regarding reference~\cite{Burin}.


\begin{thebibliography}{99}
\bibitem{footnote}
For energy states in the spin glass phase, there is a very weak decrease in $\kappa$ with $N$. This has been proposed as a feature of clustering of eigenstates in the spin-glass phase~\cite{Baldwin}.
\end{thebibliography}
\end{document}